# CONCERT: A Cloud-Based Architecture for Next-generation Cellular Systems

Jingchu Liu, Tao Zhao, Sheng Zhou, Yu Cheng, and Zhisheng Niu

# Abstract

Cellular networks are one of the corner stones of our information-driven society. However, existing cellular systems have been seriously challenged by the explosion of mobile data traffic, the emergence of machine-type communications and the flourish of mobile Internet services. In this article, we propose CONCERT (CONvergence of Cloud and cEllulaR sysTems), a converged edge infrastructure for future cellular communications and mobile computing services. The proposed architecture is constructed based on the concept of control/data (C/D) plane decoupling. The data plane includes heterogeneous physical resources such as radio interfacing equipment, computational resources, and software-defined switches. The control plane jointly coordinates physical resources to present them as virtual resources, over which software-defined services including communications, computing, and management can be deployed in a flexible manner. Moreover, we introduce new designs for physical resources placement and task scheduling, so that CONCERT can overcome the drawbacks of the existing baseband-up centralization approach and better facilitate innovations in next-generation cellular networks. These advantages are demonstrated with application examples on the radio access networks (RANs) with C/D decoupled air interface, delay sensitive machine-type communications, and real-time mobile cloud gaming. We also discuss some fundamental research issues arising with the proposed architecture to illuminate future research directions.

# 1 Introduction

Nowadays, cellular systems are faced with drastic changes and great challenges. The rapid penetration of smart phones and tablets has triggered an exponential growth of mobile data traffic in the past few years [1]. In response, both academia and industry have been devoted to enhancing the capacity of existing cellular systems with dense small cells, high-frequency bands and novel transmission technologies such as massive MIMO. It is anticipated that through these enhancements, the next-generation (5G) cellular system can have 1000x capacity and 100x transmission rate compared with 4G (LTE Release 8) systems. However, hardly have we managed to accommodate billions of mobile devices, an even greater number of machines are already waiting to communicate with each other [2]. Besides bringing a massive volume of connections, machine-to-machine (M2M) communications also have other unique communication requirements. For example, applications like fleet management and industrial control demand extremely low access latency and high reliability, which existing human-oriented cellular systems cannot satisfy. To this regard, it is also expected that 5G cellular systems can support more than 10 billion M2M communication links and down to 1ms access latency. Beyond these fast emerging requirements, cellular networks are also in danger of becoming economically unprofitable "data pipes" for the increasingly Internet-based services. To this end, 5G cellular systems should have the agility too quickly adapt to requirement changes and the capability to support a wide spectrum of new services.

Since last year, considerable effort has been devoted to investigating 5G cellular systems, and the drawbacks of the conventional cellular design paradigm have been realized. In a conventional cellular system (Figure 1a), the physical infrastructure comprises geographically distributed hardware subsystems such as base stations (BSs). These subsystems are often proprietary devices that are highly integrated and optimized for specific tasks. They communicate with each other via predefined network protocols to present network functions together. Conventional cellular systems have sufficient capabilities to serve the traffic with moderate quality of service (QoS) requirements; its subsystem-based design also facilitates modular system upgrade. However, the conventional paradigm leads to several severe problems:

- **Flexibility**: Network functions of conventional cellular systems are embodied in the proprietary subsystems and their interfacing protocols. Once these subsystems are widely deployed, it would be time-consuming and costly to upgrade functions or add services.
- **Efficiency**: The distributed system layout limits the information exchange between subsystems, and hence prohibits cooperation technologies such as coordinated multipoint transmission (CoMP) [3] and CHORUS [4], which can improve network performance with pervasive inter-cell and inter-network cooperation
- **Resource Utilization**: Geographically distributed subsystems rely on local physical resources to function. Thus, resources must be over-provisioned for peak load, leading to a low resource utilization ratio.

These problems make it hard to incorporate the features of 5G networks into conventional cellular systems.

Because of this, it is proposed that 5G networks should be constructed with centralized physical resource placement and become increasingly software-defined [5]. The baseband-up centralization architecture was proposed for radio access networks (RANs) leveraging cloud concept and technology [6-8]. In such an architecture (Figure 1b), the computational resources are aggregated from distributed sites to form a centralized computational resource pool. Meanwhile, only remote radio heads (RRHs), a kind of simple radio interfacing equipment, are left at BS sites. RRHs digitize baseband communication signals and pass them to the central pool for baseband signal processing. The benefit of baseband-up centralization is multifold: 1) the barriers against the information exchange between BSs are largely eliminated through centralization, enabling cooperation technologies; 2) computational resources can be flexibly provisioned from the pool to support regular or advanced wireless protocol processing [9]; 3) the utilization ratio of computational resources can be significantly improved through the statistical multiplexing of computational tasks [10]; 4) the centralized computational resources can be virtualized and used to support software-defined BSs, greatly simplifying the development and maintenance of cellular systems. It is worthy of noting that RAN virtualization based on baseband-up centralization is currently considered as an important use case of Network Functions Virtualization (NFV) [11], which is an industry initiative to consolidate and virtualize many network equipment types with industry standard IT servers and cloud technology.

While enlightened by previous studies, we observe some serious issues that may limit the applicability of baseband-up centralization to some promising use cases. Firstly, baseband-up centralization requires continuous exchange of raw baseband samples between RRHs and the central pool. This puts tremendous bandwidth burden[1] on the fronthaul network and may hinder the adoption of baseband-up centralization in networks with limited fronthaul resources. Secondly, all the signal processing tasks are consolidated onto the computing resource pool. As the central pool is likely to be far away from RRHs, fronthaul networking and statistical multiplexing may increase the total access latency, making it difficult to apply baseband-up centralization to some important scenarios such as extremely low latency wireless access. Last but not the least, the current design of baseband-up centralization mandates a one-to-one mapping between physical antennas and virtual BSs. Although intuitive, fixed mapping makes the implementation of some novel RAN architectures difficult, if not impossible. For example, air interface C/D decoupled RAN architectures [13-15] propose to decouple control and data channels to provide better mobility and energy saving performance for small cells. Small-coverage BSs (DBSs) which only hold data channels are densely deployed to increase system capacity, and large-coverage BSs (CBSs) which hold control channels coordinate DBSs to serve users. Implementing the C/D decoupled architecture with baseband-up centralization requires operators to provision additional radio interfacing equipment and fronthaul bandwidth for CBSs and DBSs, rather than utilize existing equipment and resources. These problems are to be addressed before baseband-up centralization can be widely adopted.

In a sense, the central pool in baseband-up centralization can be seen as a special-purpose cloud where RRHs outsource computing power. This gives an example that the cloud concept can be borrowed to enhance the performance cellular systems. Meanwhile, there have been visions that cellular systems can in turn be utilized to enhance cloud services. For example in [12], edge servers (cloudlet) are integrated into physical BSs to improve the performance of mobile multimedia services. The edge servers are in charge of data-plane multimedia processing, while the enhanced BSs control the edge servers based on session information. By reducing the distance between servers and mobile users, the latency and jitter of multimedia services can be greatly reduced. This proposal effectively extends the concept of "network edge" to the wireless edge, which is arguably the true edge in the mobile era. Nevertheless, this proposal demands new infrastructure for edge servers and thus will only further complicate the already complicated cellular infrastructure.

In this article, we propose CONCERT (CONvergence of Cloud and cEllulaR sysTems), a converged edge infrastructure for cellular networking and mobile cloud computing services. The physical resources in CONCERT are presented as virtual resources so that they can be flexibility utilized by both mobile communication and cloud computing services. The virtualization is achieved through a control/data (C/D) decoupling mechanism, by which a logical control-plane entity dynamically coordinates data-plane physical hardware such as servers, packet switches, and radio interfacing equipment in response to the requirement of services. Note that C/D decoupling has been successfully applied in SDN to orchestrate homogeneous packet-forwarding hardware, and it has also been exploited in RAN to coordinate physical BSs. Here we adopt this concept to coordinate and virtualize heterogeneous physical hardware. Our proposal provides a general framework which not only addresses the problems associated with baseband-up centralization, but also opens up new opportunities for innovations in the next-generation cellular networks: firstly, we overcome the drawbacks of baseband-up centralization by allowing flexible combination of distributed and centralized strategies in allocating data-plane computational resources for signal processing functions[2]. This gives CONCERT increased flexibility in satisfying new communication demands such as extremely low access latency and high reliability; secondly, the air interfaces of virtual BSs are decoupled from specific physical antennas and are mapped to virtual radio resources instead. This simplifies the implementation of candidate next-generation cellular architectures such as air interface C/D decoupled RANs; lastly, a common infrastructure can now be shared by mobile communication services and cloud computing services, opening up new real-time computing services for mobile users and providing new revenue sources for the operators of next-generation cellular networks.

---

[1] The rate of baseband samples varies with the modulation, bandwidth, and number of antennas used. A reference value would be 1.2Gbps for a single 20MHz LTE antenna-carrier (AxC).

The rest of the article is organized as follows. We first describe the proposed architecture with an anatomy on its physical infrastructure and software-defined services. After that, we demonstrate the advantages of CONCERT by illustrating its ability to support diverse applications. Some fundamental research issues that arise from CONCERT are then discussed to lay a foundation for future exploration. After that, we summarize our article in the conclusion.

## 2  Architecture Design

The proposed architecture is shown in Figure 1c. It contains a virtualized infrastructure and software-defined services. The infrastructure can be further divided into heterogeneous physical data-plane resources and a decoupled control plane entity called *Conductor*. The Conductor orchestrates and virtualizes data-plane resources. Software-defined services are constructed on top of virtual resources. Next, we describe each component of CONCERT in detail.

---

[2] The fully centralized placement in baseband-up centralization can now be viewed as an extreme case of the resource placement strategies in CONERT.

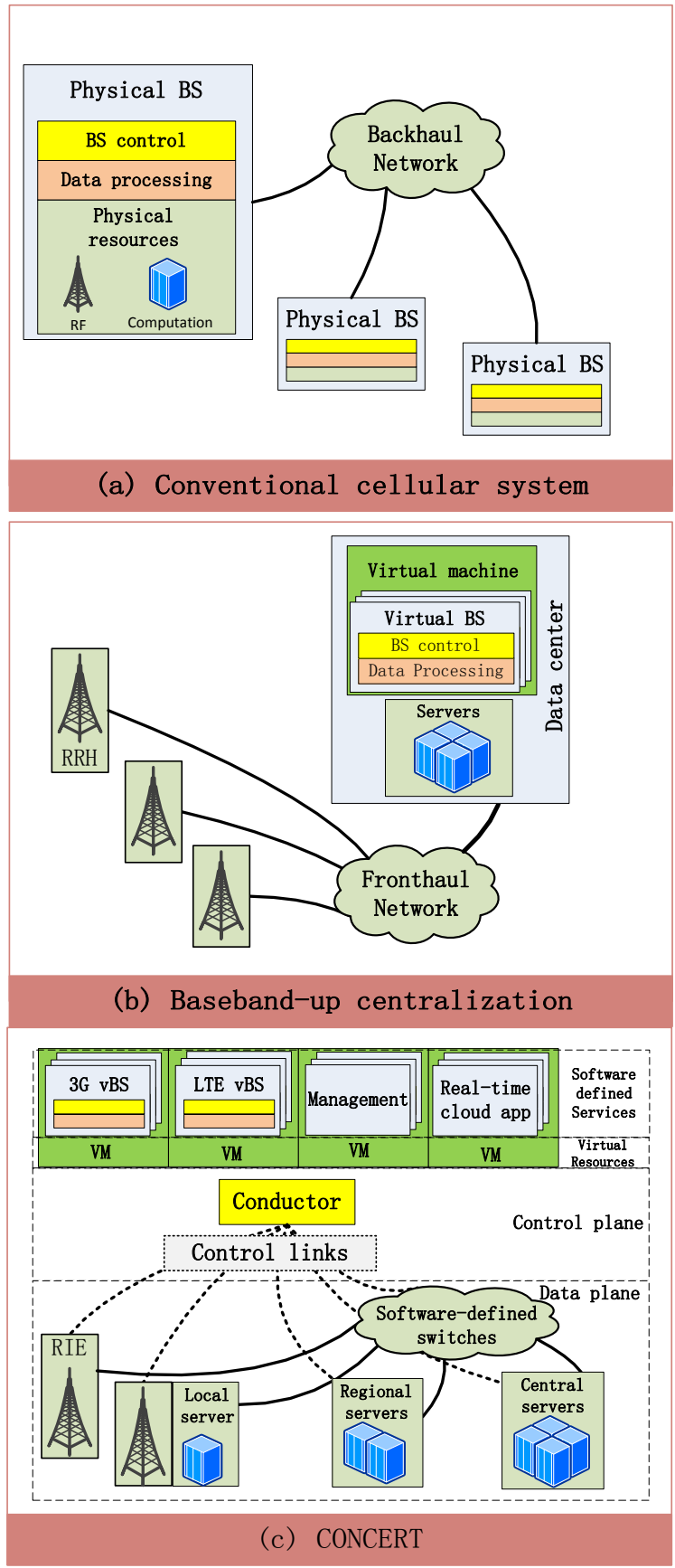

Figure 1 Comparison between conventional cellular systems, baseband-up centralization, and CONCERT architecture.

## 2.1 The Data Plane

The data plane of CONCERT consists of radio interfacing equipment, software-defined switches, and computational resources.

**Radio Interfacing Equipment (RIE):** RIEs translate signals between radio and digital domain, and perform some radio resource slicing functions. On the downlink, the RIEs convert baseband samples into radio signals and then transmit the signals via antennas. On the uplink, the RIEs receive radio signals from the air and convert them into baseband samples. Note that the RIEs differ from conventional radio frequency (RF) chains in their capabilities to reconfigure operation parameters such as the center frequencies of radio conversion and the schedules for energy saving operations. Each RIE may have different degrees of configurability, depending on implementation considerations such as form factors and equipment costs.

**Software-Defined Switches:** the RIEs and computational resources, as well as other system components, are interconnected with a software-defined transport network. The data plane of such a network is comprised of software-defined switches, which are under the control of the Conductor. Control functions such as the construction of forwarding tables are centralized in the Conductor as in SDN, so that these switches can work synergetically. Note that the switches in SDN are layer-2/3 (L2/L3) switches, but CONCERT also allows layer-0 (optical) switches such as wavelength division multiplexing (WDM) switches to provide better virtualization performance in some scenarios.

**Computational Resources:** Computational resources handle all the data-plane computation. Besides the communication-related computation such as baseband processing, they can also handle application-level computation just like the Internet cloud servers. These computational resources have different computational capabilities and are scattered in different locations: some are co-located with the RIEs to handle computational tasks with stringent latency requirements. These local computational resources need not to be very powerful, considering the limited volume of local traffic; at the same time, some computational resources possessing more computational capabilities are scattered in different places to consolidate tasks from a small region; there are also big computational resource pools with very high computational capabilities. They can consolidate large amount of computational tasks in a large area. Computational resources may contain not only general-purpose processors but also hardware accelerators, which can improve the efficiency when handling certain processing tasks.

## 2.2 The Control Plane

The Conductor coordinates physical data-plane resources on the southbound interface (towards data-plane resources) and presents them as virtual resources on the northbound interface (towards software-defined services). Although we denote the control plane as one entity, it is in fact just a logically centralized entity. The Conductor can be further designed in a hierarchical manner for better control scalability.

The southbound control functions include radio interfacing management, wired networking management and location-aware computing management. Data-plane resources receive control instructions from the Conductor and report network context information (NCI) back to it through dedicated control channels, which can be either wired (for controlling network-side resources) or wireless (for controlling user-side resources).

**Radio Interfacing Management (RIM):** The RIM module manages the radio resources to be transmitted from each RIE and adjusts the transmission power on those radio resources. Although radio resources are transmitted directly on RIEs, the RIM module may depend on some computational resources to provide flexible virtualization and fine control over radio resources. This is because the virtualization of radio resources can be more flexible with the help of software-based implementations. The RIM module can also selectively turn off some RIEs with the objective to reduce network energy consumption.

**Wired Networking Management (WNM):** The WNM module schedules the software-defined switches to provide end-to-end data delivery with QoS guarantees. It holds a database of the traffic demands from services, and also the capacities, occupancy and service policies of the switches. This database is updated frequently with the information collected. Based on this database, the WNM module instructs the switches to provision link and switching resources in order to provide desired QoS.

**Location-aware Computing Management (LCM):** The LCM module assigns computational tasks to computational resources. The computational tasks usually have various deadline requirements and computational resource demands. At the same time, the computational resources may have different computational capabilities and are usually scattered in different locations. Thus, different assignment strategies, such as full and partial centralization, will result in different system utilities. The LCM is designed to choose the assignment strategy that achieves the optimal tradeoff among multiple (possibly conflicting) objectives such as real-time requirements, resource utilization ratio, and power consumption. To do this, the LCM module needs to collect information about the deadlines, resource demands, occurrence location, and result destination of tasks. It also needs to take into account of the current occupancy of resources and the service polity of operators. Because the tasks may be sent to different physical locations, the assignment strategy is also affected by the QoS of the virtual links.

On the northbound interface, all the physical resources are virtualized, forming a virtual infrastructure. There are three types of virtual resources in this virtual infrastructure.

**Virtual Computational Resources:** Computational resources can be presented in the form of virtual machines (VMs). These VMs are different from the VMs in the Internet clouds because they should be provisioned to meet real-time processing requirements. To achieve this, the Conductor calls the LCM module to tune the capacity and location of the corresponding physical computational resources, and to schedule computational tasks. These VMs may be dynamically migrated to different physical resources in response to time-varying environments, so that the performance of VM can be maintained.

**Virtual Networking Resources:** Virtual networking resources are presented in the form of end-to-end links with certain QoS guarantees such as bandwidth and delay. These guarantees can be vastly different for different kinds of services. For instance, the baseband links used in communication services may need Gbps-level bandwidth and 100us-level delay, while the links used in mobile cloud services may only need Mbps-level bandwidth and 10ms-level delay. The Conductor relies on the WNM module and heterogeneous data-plane networking resources to provision virtual links with such diverse guarantees.

**Virtual Radio Resources:** The Conductor virtualizes radio resources with the help of the RIM

module. Virtual radio resources are presented in the form of time-frequency blocks transmitted or received at certain locations with some signal-to-noise-and-interference ratio (SINR) guarantees. We take such a low level virtualization because it is more tractable to build communication systems on top of low level radio resources. Of course, higher-level virtualization may be easier to use when constructing user applications, but this may complicate the design of CONCERT and degrade its performance. So we leave the task of constructing higher-level virtual radio resources for the software-defined services.

## 2.3 Software-Defined Services

On top of the virtual resources, various services can be conveniently deployed in a software-defined manner. Instead of directly handling the physical hardware, service developers now only have to manage the on-demand and easy-to-use virtual resources. This can greatly simplify the process of service construction and deployment. A basic service that COCNERT can support is virtual base stations (vBSs). The operators just have to initiate an instance of vBS software and the Conductor will provision the required virtual resources so that the functions of a physical BS can be delivered. Traditional mobile communication services can be further realized on top of those vBSs. CONCERT also supports the deployment of novel cellular systems. For example, to deploy a baseband-up centralization RAN, the Conductor can provision radio resources for some RIEs to form virtual antennas; virtualize a portion of the underlying software-defined network as the fronthaul network; and virtualize some computational resources in a data center as the baseband processing pool. Beyond communication systems, mobile cloud computing services can also be deployed and provided for mobile users. A service provider just needs to require some virtual machines and some wired links from CONCERT, and then a mobile cloud computing service with delay guarantees can be constructed thereon with the mechanisms used in Internet Cloud.

# 3 Application Examples

With a fully virtualized infrastructure, CONCERT can support the construction and deployment of a wide variety of software-defined services. Here we present three application examples to demonstrate the capability of CONCERT.

## 3.1 Hyper-Cellular Networks

Small cell is recognized as a key mechanism to increase the capacity of 5G cellular systems. Yet, the shrinking cell size and the increasing node number pose some mobility and energy consumption problems. To overcome these problems, a number of RAN architectures with air interface C/D decoupling have been proposed. The hyper-cellular network (HCN) [13] is designed to improve the energy efficiency and radio resource utilization by splitting the air interface into two layers: an always-on control layer and an on-demand data layer[3]. In implementation, the BSs in HCN are correspondingly categorized into two types: wide-coverage control base stations (CBSs) which serve the control layer, and small-coverage data base stations (DBSs) which serve the data layer. A CBS coordinates several DBSs to server the mobile users within its coverage. The decoupled control coverage and centralized coordination of CBSs can effectively transform the nature of mobility management from inter-cell handover in traditional small cell settings to intra-cell scheduling. Also, the energy consumption can be reduced by selectively turning off lightly-loaded DBSs.

With conventional RAN infrastructures, it is difficult to implement the decoupled air interface and centralized control of HCN. However the implementation of HCN becomes convenient in CONCERT. As illustrated in Figure 2, to construct a HCN, CBSs and DBSs are initiated as software instances running in virtual machines provisioned in a data center and are mapped with virtual radio resources at indicated RIE sites. Those two kinds of vBSs are connected with each other via high-speed virtual links inside the data center. In this way, the CBSs and DBSs can efficiently exchange information so that CBSs can perform centralized control functions such as offloading, handover and cell on-off. To differentiate the coverage areas of the CBSs and DBSs, the Conductor will assign different transmission power on provisioned virtual radio resources. If a DBS is turned into sleep mode, the Conductor can reclaim the radio resources assigned to that DBS temporally for other usage. In addition, other systems such as LTE can coexist with HCN on the same infrastructure as long as their VBS instances are held in separate VMs and they are assigned with physically orthogonal virtual radio resources.

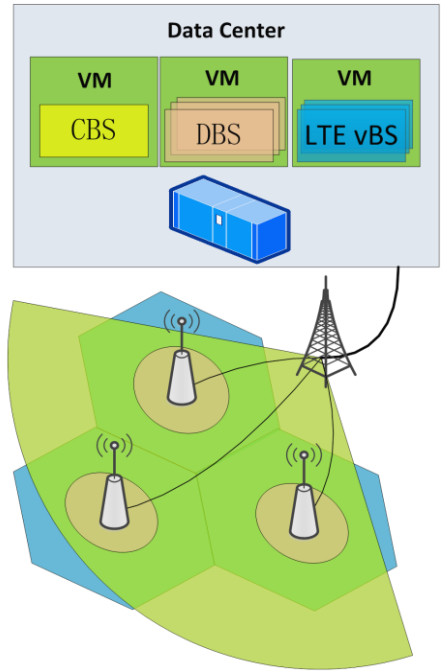

Figure 2 Implementation of HCN in CONCERT

---

[3] Improper decoupling of the air interface may cause performance penalties. For this reason, the control channels that are tightly coupled with data channels need to be kept together with these data channels; on the other hand, the control channels that are not tightly-coupled with data channels can be decoupled from the antennas that these data channels reside and be placed on a separate antenna. We refer interested readers to [13-15] for more details about the implementation of air interface C/D decoupling.

## 3.2 Delay and Reliability Sensitive Networking

CONCERT can better meet the unique requirements of M2M communications compared with conventional cellular systems for the following reasons: 1) radio resources can be flexibly allocated and scheduled to support a larger number of connections; 2) baseband computational tasks can be executed at the locations with rich computational resources so that advanced decoding and/or multiuser detection can be conducted to improve communication reliability and support more connections; 3) localized computing and QoS-guaranteed networking can reduce the latency of task execution and data delivery, enabling delay sensitive networking applications.

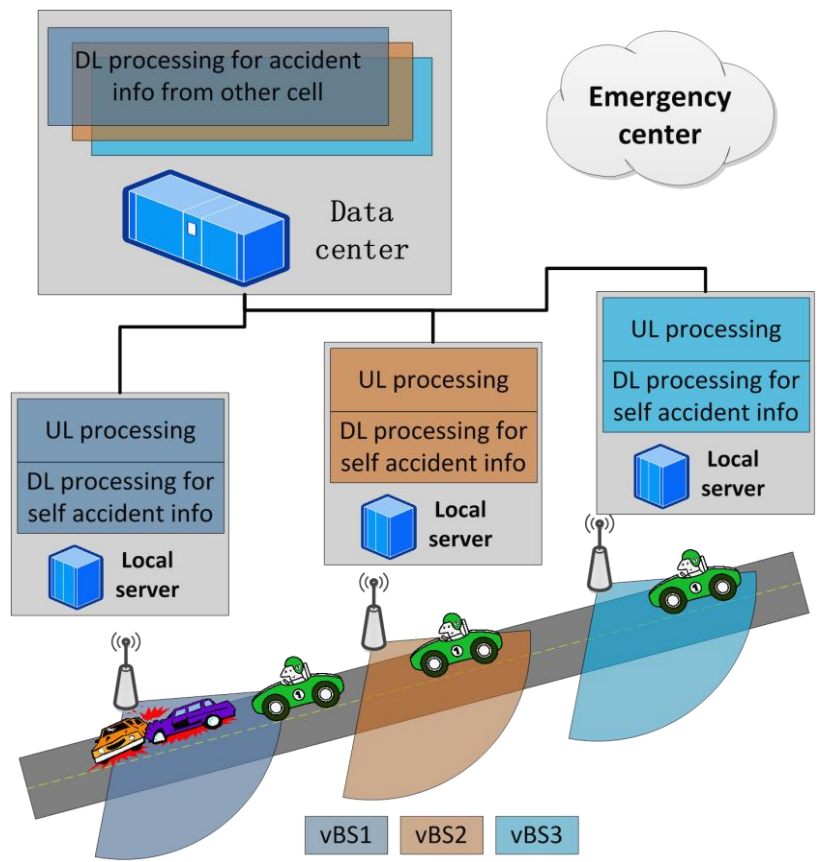

Figure 3 Implementation of a vehicle accident report system in CONCERT

Here we give an example on vehicle accident report system to show how CONCERT can support delay and reliability sensitive M2M communications. As illustrated in Figure 3, the system consists of several vBSs that cover a highway. For each vBS, the Conductor provisions RIEs, local computational resources alongside RIEs and centralized computational resources in a data center. The Conductor also provisions virtual wired networking resources to deliver data among the RIEs, computational resources, and the Internet. Suppose that an accident happens in the coverage area of vBS1, the embedded accident detection systems on the involved vehicle will automatically report to vBS1 via a dedicated uplink channel until a response is received. To decode the accident information quickly, vBS1 allocates baseband processing tasks to local computational resources. If the decoding is successful, vBS1 will instantly generate broadcast frames and send them out on the downlink so that all vehicles nearby can get warned and slow down. The decoded information will also be sent to other vBSs to inform vehicles in faraway cells. The downlink baseband processing of other vBSs is executed on the centralized computational resources to save the local computational resources for the processing of other accident information. Moreover, accident information can be sent to an emergency center to facilitate timely rescue.

## 3.3 Real-Time Cloud Computing Service

Mobile phone users enjoy cloud services by which computational and storage capabilities can be enhanced at low costs. Take cloud gaming service as an example: game engines require massive computation to render 3D animation and to simulate physical interaction. Hence operating game engines will incur high power consumption. As a result, current mobile devices usually do not support 3D video gaming well. However, if the cloud gaming service is introduced, mobile devices can offload most of the computation to the server applications hosted in clouds to save energy and boost user experience. Nevertheless, data delivery across wide area networks (WAN) may experience long latency. When providing cloud gaming services, the server-side game engine instance needs to interact with the user within a delay constraint beyond what current Internet Clouds (ICs) can provide. Consequently, delay sensitive applications such as computation offloading and interactive cloud gaming are often poorly supported by ICs.

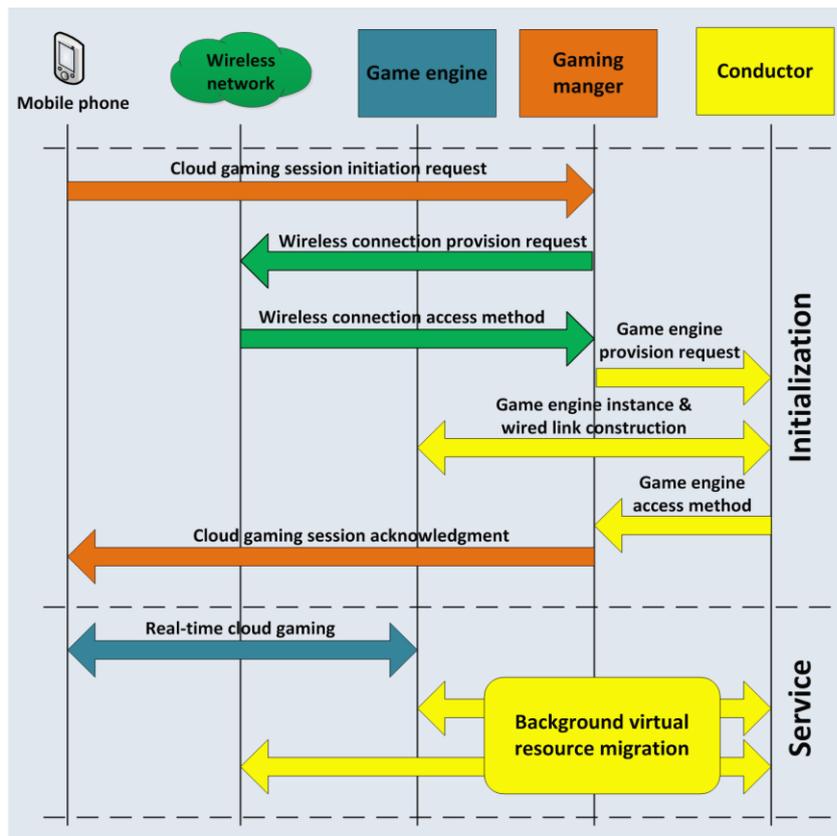

Figure 4 Signaling procedure of a cloud gaming service implemented in CONCERT

In contrast, CONCERT can provide computational resources located much closer to the users than those of the ICs. Furthermore, the latency of data delivery within CONCERT can be guaranteed through proper allocation of switching resources. In this way, real-time cloud gaming service constructed. Specifically, to construct a cloud gaming service, the service provider first creates a gaming manager (GM) application on the virtual machine. As illustrated in Figure 4, a user can initiate a service session by sending a request to the GM using any kind of mobile data services. In response, the GM negotiates with a wireless network deployed in CONCERT for virtual wireless links to connect with the requesting user. At the same time, the GM also require the Conductor for another virtual machine to run a game engine instance and a virtual wired link as a real-time connection between the game engine instance and the serving wireless network. The Conductor will provision virtual resources accordingly. After the required resources are properly configured, the GM will acknowledge the user with the method to access the wireless network and the game engine, so that an end-to-end link between the user and the game engine can be established. The user and the game engine then start to communicate. On the downlink direction, the game engine generate the video frames to be displayed on the user's screen, encodes them and sends them to the user. The user receives and decodes the data, and then displays the frames. On the uplink direction, the user reports his/her input to the game engine, which then decides the forthcoming frames accordingly. When the user and the game engine are communicating, the Conductor monitors the mobility of the user through the NCI updates. If the physical resources in use no longer satisfy the negotiated QoS, the Conductor will migrate the functionalities involved, such as the virtual machine running the game engine instance, to more proper physical resources so as to maintain the negotiated QoS.

# 4  Fundamental Research Issues

As discussed above, CONCERT is capable of supporting a wide variety of services through the virtualization of data-plane physical resources. Such a cloud-based design poses some fundamental issues that demand further investigation.

## 4.1  Mathematical Modeling and Analysis

Proper resource placement and task scheduling are the basis for providing desired QoS in CONCERT. However, the optimal resource placement and efficient task scheduling are challenging research issues. To get the best placement and scheduling strategy, we may need to solve optimization problems with multiple (possibly conflicting) objectives such as real-time requirements, resource utilization, and power consumption. Different decisions on placement and scheduling strategies will achieve different tradeoffs between those objectives. Take centralized and localized physical layer (PHY) processing as an example: through centralized processing, we can reduce the processing latency by provisioning more computational resources, but aggregating tasks from distributed places will inevitably incur some networking delay. In contrast, localized processing may consume more processing time because the computational resources are more limited, but the networking delay can be eliminated. The optimal strategies that achieve minimum overall latency can only be obtained after carefully evaluating the tradeoff between the processing and networking latency. Moreover, the optimal strategy will become more complicated when we take other objectives, such as resource utilization and bandwidth requirements, into consideration.

To address these problems, an analytical framework based on queueing theory may be developed. With such a framework, computational tasks can be modeled as customers, and resources (e.g. local computational resources, processors in the centralized pool, switches) can be modeled as queuing nodes. Customers travel across the network of queues according to some routing policy and are serviced at each node according to some scheduling policy enforced by the Conductor. Within this framework, various performance metrics can be explicitly evaluated. For example, the task processing latency can be evaluated as the total time the corresponding customer spends traveling from the source node to the destination node, and resources utilization can be mapped to the utilization of corresponding nodes. These results will provide a mathematical foundation for optimizing or finding tradeoff between objectives.

## 4.2 Control and Data Plane Realization

The control and data planes of CONCERT consist of many components that demand novel design and realization. One example is the Conductor. As it controls a large number of data-plane resources, realizing it using a physically centralized scheme may bring scalability problems. As a solution, the Conductor can be realized in a hierarchical manner: only those control functions that affect global performance are physical centralized; those control functions that merely affect local network performance can be handled by a local control entity; additionally, when control- and data-plane functions are tightly coupled, these functions must be handled by a local controller to avoid performance penalties. While the hierarchical control-plane structure may be intuitive, the optimal way to realize it is still an open problem.

The realization of RIE is yet another challenging issue. The RIE is designed reconfigurable to flexibly accommodate the requirements of different scenario. For example in HCN, the control and data radio channels can be configured to transmit on different RIEs and with different transmission power. Also, more bandwidth can be allocated for random access channels in delay sensitive communications so as to reduce collision probability and minimize access latency. This re-configurability makes RIE distinct from the existing simple radio interfacing equipment and may require more innovative design to strikes a balance between efficiency and flexibility.

## 4.3 Resource Virtualization and Service Provisioning

Resource virtualization is fundamental to the deployment of software-defined services, but the heterogeneous nature of CONCERT's data plane poses several difficulties on resource virtualization. First of all, the features and constraints of physical resources should be taken into account in virtualization, and proper physical resources should be utilized to guarantee virtualization efficiency. For example, virtualizing low-latency fronthaul links using the WDM networks is much more efficient and straightforward than using packet-switched networks such as Ethernet. The same argument applies when virtualizing real-time VMs for VBS software: the virtualization efficiency would be much higher using hardware accelerators instead of a pure general-purpose platform. Secondly, sometimes heterogeneous resources need to be orchestrated to guarantee virtualization efficiency. Take delay sensitive networking as an example. As the baseband processing functions are distributed between local and central sites in this example, the VMs holding the vBSs can only be virtualized through the synergetic coordination of networking and computational resources. Lastly, the virtual resources should be presented in easy-to-use forms. When building a user application, virtual links with guaranteed throughput and/or delay is more understandable than those with guarantees on SINR. The situation is on the contrary for baseband processing, where the performance of processing algorithms relies directly on SINR levels.

Moreover, a user- and developer-friendly interface should be constructed to facilitate the process of service construction and utilization. This interface is expected to incorporate functions such as service request, authentication, authorization, billing as well as virtual resource utilization and service construction. Also, the services deployed in CONCERT are preferred to be interoperable with services provided by other ecosystems to build more powerful applications. A very straightforward example would be the interplay with Internet Cloud (IC). CONCERT can handle delay-sensitive computations while ICs can take care of delay-tolerant tasks. In this way, a hybrid cloud computing service can be efficiently provided leveraging both the real-time feature of CONCERT and the economy of scale in IC. Other forms of ecosystem interplay call for further research.

# 5  Conclusion

In this article, we propose a converged edge infrastructure for future mobile communication and cloud computing services. We refactored the conventional cellular infrastructure with decoupled control plane and heterogeneous reconfigurable data plane. The control plane serves to coordinate and virtualize physical data-plane resources. With the virtual resources, a wide variety of software-defined services ranging from mobile communications to real-time cloud computing can be supported. CONCERT avoids the drawbacks of baseband-up centralization by combining centralized and distributed strategies in resource placement and utilizing virtual radio resources. It also enables the deployment of real-time mobile cloud computing services. We illustrate the capability of CONCERT with three novel use cases and also discuss some fundamental research issues related.

# Acknowledgment

This work is sponsored in part by the National Basic Research Program of China (973 Program: 2012CB316001), the National Science Foundation of China (NSFC) under grant No. 61201191, the Creative Research Groups of NSFC under grant No. 61321061, and Intel Corporation.

# 6  Biographies

JINGCHU LIU (liu-jc12@mails.tsinghua.edu.cn) received his B.S. degree in Electronic Engineering from Tsinghua University, China, in 2012. He is currently a PhD student at the Department of Electronic Engineering, Tsinghua University. His research interests include cloud-based cellular/wireless networks, cellular big data, and green wireless communications.

TAO ZHAO (t-zhao12@mails.tsinghua.edu.cn) received his B.S. degree in Electronic Engineering from Tsinghua University, China, in 2012. He is currently a master student in Department of Electronic Engineering, Tsinghua University. His current research interests include green wireless communications, software defined radio, and cellular network architectures.

SHENG ZHOU (sheng.zhou@tsinghua.edu.cn) received his B.S. and Ph.D. degrees in Electronic Engineering from Tsinghua University, China, in 2005 and 2011, respectively. He is currently an assistant professor of the Department of Electronic Engineering, Tsinghua University. From January to June 2010, he was a visiting student at Wireless System Lab, Electrical Engineering Department, Stanford University, CA, USA. His research interests include cross-layer design for multiple antenna systems, cooperative transmission in cellular systems, and green wireless cellular communications.

YU CHENG (cheng@iit.edu) received the B.E. and M.E. degrees in Electronic Engineering from Tsinghua University, Beijing, China, in 1995 and 1998, respectively, and the Ph.D. degree in Electrical and Computer Engineering from the University of Waterloo, Canada, in 2003. He is now an Associate Professor with the Department of Electrical and Computer Engineering, Illinois Institute of Technology, USA. His research interests include next-generation Internet architectures and management, wireless network performance analysis, network security, and wireless/wireline interworking. He received a Best Paper Award from the conferences QShine 2007, IEEE ICC 2011, and a Runner-Up Best Paper Award from ACM MobiHoc 2014. He received the National Science Foundation (NSF) CAREER AWARD in 2011 and IIT Sigma Xi Research Award in the junior faculty division in 2013. He served several Symposium Co-Chairs in IEEE ICC and IEEE GLOBECOM, a Technical Program Committee (TPC) Co-Chair for WASA 2011 and ICNC 2015. He is a founding Vice Chair of the IEEE ComSoc Technical Subcommittee on Green Communications and Computing. He is an Associated Editor for IEEE Transactions on Vehicular Technology and the New Books & Multimedia Column Editor for IEEE Network. He is a senior member of the IEEE.

ZHISHENG NIU (niuzhs@mail.tsinghua.edu.cn) graduated from Beijing Jiaotong University, China, in 1985, and got his M.E. and D.E. degrees from Toyohashi University of Technology, Japan, in 1989 and 1992, respectively. During 1992-94, he worked for Fujitsu Laboratories Ltd., Japan, and in 1994 joined with Tsinghua University, Beijing, China, where he is now a professor at the Department of Electronic Engineering and deputy dean of the School of Information Science and Technology.  He is also a guest chair professor of Shandong University, China.  His major research interests include queueing theory, traffic engineering, mobile Internet, radio resource management of wireless networks, and green communication and networks. He is now a fellow of both IEEE and IEICE, a distinguished lecturer (2012-15) and Chair of Emerging Technology Committee (2014-15) of IEEE Communication Society, and a distinguished lecturer (2014-16) of IEEE Vehicular Technologies Society.